\documentclass[reprint,
preprintnumbers,superscriptaddress,endnote,nofootinbib,
aps,prd,twocolumn,
floatfix,
10pt]{revtex4-2}
\usepackage{amssymb,amsmath}
\usepackage[T1]{fontenc}
\usepackage[utf8]{inputenc}
\usepackage{wrapfig}
\usepackage{graphicx}
\usepackage{dcolumn}
\usepackage{bm}
\usepackage{float}
\usepackage[caption=false]{subfig}
\usepackage[colorlinks=true,allcolors=blue]{hyperref}

\usepackage{placeins}

\begin{document}
\title{Cosmic Strings, Inflation, and Gravity Waves}	
\author{George Lazarides}
\affiliation{School of Electrical and Computer Engineering, Faculty of Engineering, Aristotle University of Thessaloniki, Thessaloniki 54124, Greece}	
\author{Rinku Maji}
\affiliation{Department of Physics, Indian Institute of Technology Kanpur, Kanpur-208016, India}
\author{Qaisar Shafi}
\affiliation{Bartol Research Institute, Department of Physics and Astronomy,
	University of Delaware, Newark, DE 19716, USA}

\begin{abstract}
We investigate the impact of Coleman-Weinberg inflation on the 
stochastic gravity wave background spectrum emitted by intermediate scale cosmic strings. The string network is partially inflated and re-enters the horizon at later times after the end of inflation, such that the short string loops are not produced. This leads to a significant modification of the gravity wave spectrum that we explore  in detail.
We find that Coleman-Weinberg inflation can help to satisfy the Parkes Pulsar Timing Array (PPTA) bound for dimensionless string tension values in the range $G\mu > 1.1\times 10^{-10}$. We also identify the modified gravity wave spectra which, in the case of inflation, are compatible with the North American Nanohertz Observatory for Gravitational Waves (NANOGrav) data. We then discuss the formation of monopoles and strings at the same breaking scale and the compatibility of the Monopole, Astrophysics and Cosmic Ray Observatory (MACRO) bound with the PPTA bound, and also with the NANOGrav data. Finally, an example of a realistic non-supersymmetric $E_6$ model
incorporating successful Coleman-Weinberg inflation is presented in which monopoles and strings both survive inflation and are present at an 
observable level.
\end{abstract}	
\maketitle
\flushbottom

\section{Introduction}
The existence of topologically stable cosmic strings in grand unified theory (GUT) models such as 
$SO(10)$ (more precisely $Spin(10)$) has been known for quite some time \cite{z2string}, and their role in cosmology has attracted a fair amount of interest in recent years -- see e.g. Refs.~\cite{TopDef,ellis,cui,Chakrabortty:2020otp,cosmicstrings1,cosmicstrings2}. The string tension $\mu$, i.e. the energy per unit length of the string, depends on the symmetry breaking pattern of the GUT symmetry and is related to an appropriate intermediate scale $M_{\mathrm{str}}$ determined by various phenomenological considerations including gauge coupling unification. The primordial string loops decay by emitting stochastic gravitational radiation \cite{vilenkinbook}, and an important constraint on $M_{\mathrm{str}}$ arises from the pulsar timing array  experiments -- see Ref.~\cite{pta} and references therein. The recently published 12.5 yr pulsar timing array data from the NANOGrav collaboration \cite{NANOGrav} provides some provisional evidence for the existence of a gravity wave signal at frequencies 
$f\sim 1~{\rm yr}^{-1}$. It has been recognized by several authors \cite{ellis,NANOGrav,cosmicstrings2} that these data appear compatible with an interpretation in terms of a stochastic gravity wave background emitted by cosmic strings.
As argued in the NANOGrav paper \cite{NANOGrav}, the apparent tension between their results and the PPTA bound \cite{ppta} is not real, but comes from the use of the improved prior for the pulsar red noise \cite{Hazboun}. Therefore, we either take the NANOGrav result as a true detection of the stochastic gravity waves and abandon the PPTA limit, or conversely, we consider the PPTA limit as a genuine bound and discard the NANOGrav signal.

In this paper our major motivation is to explore the impact of inflation experienced by strings on the gravity wave spectrum. The idea is that the strings are generated in a phase transition during primordial inflation. They are then partially inflated and at some point after the end of inflation, they re-enter the  post-inflationary horizon. Their subsequent self interactions produce string loops which emit gravity waves and eventually decay. Under these circumstances, the relatively short loops are absent and, consequently, the gravity wave spectrum at high frequencies is altered. Indeed, as the re-entrance time of the strings approaches the equidensity time, the spectrum of gravity waves at frequencies corresponding to its overall peak or higher is gradually reduced \cite{Chakrabortty:2020otp}. This scenario produces a modified spectrum which, if necessary, can help comply with the PPTA bound for appropriate $G\mu$ values ($G\mu$ is the
dimensionless string tension with $G$ being Newton's constant and $\mu$
the mass per unit length of the string). In this paper we assume that the gravity waves arise solely from cosmic strings.
  
It is interesting to note that $G\mu$ values of order $10^{-11}$ or so, which are of interest here, require intermediate scales $M_{\mathrm{str}} \sim 10^{14}$ GeV which coincide, more or less, with the values of the Hubble parameter during primordial inflation driven by a gauge singlet real scalar field with a Coleman-Weinberg potential \cite{vilenkin,plb148}. This fact has been exploited in the past to show how intermediate scale primordial monopoles may survive inflation and appear in the present universe at an observable rate \cite{Chakrabortty:2020otp,plb148,nefer}. By the same token, intermediate scale cosmic strings can also survive an inflationary epoch and contribute to the present spectrum of stochastic gravity waves. Therefore, it is instructive to consider this particular inflationary model and study its impact on the gravity wave spectra in the light of the PPTA bound or NANOGrav data.

The paper is organized as follows. In Sec.~\ref{sec:inf}, following closely the discussion in 
Ref.~\cite{Chakrabortty:2020otp}, we summarize the salient features of the inflationary model with a Coleman-Weinberg potential and sketch briefly the intermediate scale phase transition leading to string formation. In Sec.~\ref{sec:strings} we outline the evolution of the string network and the calculation of the gravity wave spectrum that is generated. Sec.~\ref{sec:tension} shows how inflation with the Coleman-Weinberg potential for a real GUT singlet modifies the gravity wave spectra from the intermediate mass strings, such that it can satisfy the PPTA bound on the string tension. In Sec.~\ref{sec:power} we fit the power-law approximation of the gravity wave spectrum for various $G\mu$ values both with and without inflation. In the case of inflation, we find the modified gravity wave spectra from strings which are compatible with the NANOGrav data. In Sec.~\ref{sec:monopole_string} we discuss the formation of monopoles and strings at the same intermediate breaking scale and the compatibility of the MACRO bound \cite{macro} with the PPTA bound, and also with the NANOGrav data. We present a realistic non-supersymmetric $E_6$ model which incorporates successful inflation with a Coleman-Weinberg potential. Our results are summarized in Sec.~\ref{sec:concl}.    
\section{Inflation and Phase Transition}
\label{sec:inf}
We employ an inflationary scenario where the inflaton is a gauge singlet real scalar field $\phi$ with 
a Coleman-Weinberg potential \cite{vilenkin,plb148}
\begin{equation}\label{CW-potential}
V(\phi)= A\phi^4\left[ \log\left(\frac{\phi}{M}\right) - \frac{1}{4}\right] +V_0.
\end{equation}
Here $V_0=AM^4/4$, $M$ is the vacuum expectation value (VEV) of $\phi$, and $A = \beta^4 D/(16\pi^2)$ \cite{TopDef}, where $D$ is the dimensionality of the representation to which the GUT gauge symmetry breaking real scalar field 
$\chi$ belongs, and $\beta$ determines the coupling $-\beta^2\phi^2\chi^2/2$ between $\phi$ and 
$\chi$. For definiteness, we adopt the particular parameter set $V_0^{1/4} = 1.66\times 10^{16}$ GeV, 
$M = 23.81\,m_{\mathrm{Pl}}$, and $A = 2.5\times 10^{-14}$ from Table 6 of Ref.~\cite{Chakrabortty:2020otp} corresponding to a viable inflationary scenario ($m_{\rm Pl} = 2.4\times 10^{18}$ GeV is the reduced Planck mass). The inflaton value at the pivot scale $k_* = 0.05 \ \mathrm{Mpc^{-1}}$ and at the end of inflation is $\phi_*=12.17\, m_{\mathrm{Pl}}$ and $\phi_e = 22.47\, m_{\mathrm{Pl}}$ respectively. The termination of inflation is determined by the condition max($\epsilon$,$|\eta|$)=1, where $\epsilon$ and $\eta$ are the usual slow-roll parameters (for a review see Ref.~\cite{lythbook}).
 
We assume that the cosmic strings are generated during a phase transition associated with an intermediate step of gauge symmetry breaking by the VEV of the real scalar field $\chi_{\mathrm{str}}$. This is the canonically normalized real component of a scalar field belonging to a nontrivial representation of the gauge group. The $\chi_{\mathrm{str}}$-dependent part of the potential is
\begin{equation}\label{potential_chi1}
V(\phi,\chi_{\mathrm{str}})=-\frac{1}{2}\beta_{\mathrm{str}}^2 \phi^2 \chi_{\mathrm{str}}^2 + \frac{\alpha_{\mathrm{str}}}{4} \chi_{\mathrm{str}}^4,
\end{equation}
implying that the VEV of $\chi_{\mathrm{str}}$ after the end of inflation is given by
\begin{equation}
\langle\chi_{\mathrm{str}}\rangle \equiv M_{\mathrm{str}}= \frac{\beta_{\mathrm{str}}}{\sqrt{\alpha_{\mathrm{str}}}} M .
\end{equation}
During inflation, the finite temperature corrections to the potential in 
Eq.~(\ref{potential_chi1}) contribute an additional term $(1/2)\sigma_{\chi_{\mathrm{str}}}T_H^2\chi_{\mathrm{str}}^2$, where $\sigma_{\chi_{\mathrm{str}}}$is of order unity and 
$T_H=H/2\pi$ is the Hawking temperature, with $H$ being the Hubble parameter. Two symmetric minima of the potential appear at 
\begin{align}\label{min}
\chi_{\mathrm{str}}= \pm \sqrt{\left(\beta_{\mathrm{str}}^2 \phi^2 - \sigma_{\chi_{\mathrm{str}}}T_H^2\right)/\alpha_{\mathrm{str}}}
\end{align}
as $\phi$ grows sufficiently large. The effective mass $m_{\mathrm{eff}}$ of 
$\chi_{\mathrm{str}}$ at these minima is given by
\begin{equation}
\label{effective-mass_chi1}
m_{\mathrm{eff}}^{2}=2\left(\beta^2_{\rm str}\phi^2-\sigma_{\chi_{\mathrm{str}}} T_H^2\right).
\end{equation}

The phase transition during which the intermediate gauge symmetry breaking is completed and the 
strings are formed occurs when the Ginzburg criterion \cite{ginzburg}
\begin{equation}
\frac{4\pi}{3}\xi^{3}\Delta V=T_H
\label{criterion}
\end{equation}
is satisfied (for details see Ref.~\cite{Chakrabortty:2020otp}). Here 
\begin{equation}
\Delta V=\frac{1}{4\alpha_{\rm str}}\left(\beta_{\rm str}^2\phi^2-
\sigma_{\chi_{\rm str}}T_H^2\right)^2=\frac{m_{\rm eff}^4}{16\alpha_{\rm str}}
\label{deltaV}
\end{equation}
is the potential energy difference between the minima in Eq.~(\ref{min}) and the local maximum at 
$\chi_{\rm str}=0$, 
and 
\begin{equation}
\xi={\rm min}\left(H^{-1},m_{\rm eff}^{-1}\right)
\label{correlation}
\end{equation} 
is the correlation length. Using Eqs.~(\ref{criterion}), (\ref{deltaV}), and (\ref{correlation}), one can show that $m_{\rm eff}^{-1} \leq H^{-1}$ implies that $\alpha_{\rm str}\geq \pi^2/6$ and 
$m_{\rm eff}^{-1} \geq H^{-1}$ implies that $\alpha_{\rm str}\leq \pi^2/6$. In the former case, 
$\xi=m_{\rm eff}^{-1}$ and the Ginzburg criterion in Eq.~(\ref{criterion}) takes the form 
\begin{equation} \label{form_mon}
\beta_{\mathrm{str}}^2\phi^2=\left( \frac{72\alpha_{\mathrm{str}}^2}{\pi^2}+\sigma_{\chi_{\mathrm{str}}}\right) T_H^2. 
\end{equation}
The intermediate symmetry breaking scale can then be expressed as
\begin{equation}\label{breaking_scale_1}
M_{\mathrm{str}} = \sqrt{\left(\frac{72\alpha_{\mathrm{str}}^2}{\pi^2}+\sigma_{\chi_{\mathrm{str}}} \right)}\, \frac{H_{\mathrm{str}}}{2\pi \phi_{\mathrm{str}}}\, \frac{M}{\sqrt{\alpha_{\mathrm{str}}}},
\end{equation}
where $\phi_{\mathrm{str}}$ is the inflaton field value at the phase transition, 
$H_{\mathrm{str}}=\sqrt{V(\phi_{\rm str})/3m_{\rm Pl}^2}$ is the corresponding value of the Hubble parameter, and we set $\sigma_{\chi_{\mathrm{str}}}=1$. On the other hand, for $\alpha_{\rm str}\leq 
\pi^2/6$, Eq.~(\ref{criterion}) gives
\begin{equation}\label{breaking_scale_2}
M_{\mathrm{str}} = \sqrt{\left(2 \sqrt{6}\pi\alpha_{\mathrm{str}}^{\frac{1}{2}}+\sigma_{\chi_{\mathrm{str}}}\right)}\, \frac{H_{\mathrm{str}}}{2\pi \phi_{\mathrm{str}}}\, \frac{M}{\sqrt{\alpha_{\mathrm{str}}}}.
\end{equation}
\section{Cosmic Strings and Gravity Waves}
\label{sec:strings}
The dimensionless tension of the cosmic strings formed during the symmetry breaking at the intermediate scale $M_{\mathrm{str}}$ is \cite{stringtension} 
\begin{equation}\label{Gmu}
G\mu\simeq\frac{1}{8}B(\frac{\alpha_{\mathrm{str}}}{g^2})\left(\frac{M_{\rm str}}{m_{\rm Pl}}\right)^2 ,
\end{equation} 
where $g=g_U\sqrt{2}$ is the relevant effective gauge coupling constant, with $g_U\simeq 0.5$ being the unified gauge coupling, and the function 
\begin{align}\label{Bx}
B(x)= \begin{cases} 
1.04 \ x^{0.195} & \mbox{for} \ 10^{-2}\lesssim x\lesssim 10^2 \\
2.4/\ln(2/x) & \mbox{for} \ x\lesssim 0.01.
\end{cases}
\end{align}
The mean inter-string separation at the time of formation is $p~\xi(\phi_{\rm str})$, where $p\simeq 2$ is a geometric factor \cite{TopDef,Chakrabortty:2020otp}. This distance increases by a factor $\exp(N_{\rm str})$ during inflation and by
$(t_r/\tau)^{2/3}$ during inflaton oscillations, where $N_{\mathrm{str}}=(1/m_{\rm Pl}^2)
\int_{\phi_e}^{\phi_{\mathrm{str}}} V \mathrm{d}\phi/V^\prime$ is the number of $e$-foldings experienced by the strings, $t_r$ is the reheat time, and $\tau$ is the time when inflation ends.

The mean inter-string distance at cosmic time $t$ after reheating is given by
\begin{equation}\label{inter-string-dist}
d_{\rm str}=p~\xi(\phi_{\mathrm{str}})\exp(N_{\mathrm{str}})\left(\frac{t_r}{\tau}\right)^{\frac{2}{3}} 
\frac{T_r}{T}, 
\end{equation}
with $T_r$ being the reheat temperature, the temperature $T$ during radiation dominance given by
\begin{equation}\label{time-rad-dom}
T^2=\sqrt{\frac{45}{2\pi^2g_*}}\,\frac{m_{\rm Pl}}{t},
\end{equation}
and $g_*$ accounting for the appropriate value of the effective number of massless degrees of freedom  for the relevant temperature range. For the numerical example considered in Sec.~\ref{sec:inf}, 
$T_r=10^9~{\rm GeV}$, $t_r\simeq 0.36~{\rm GeV}^{-1}\simeq 2.37\times 10^{-25}~{\rm sec}$, and 
$\tau\simeq 1.26\times 10^{-12}~{\rm GeV}^{-1}\simeq 8.3\times 10^{-37}~{\rm sec}$. The string network 
re-enters the post-inflationary horizon at cosmic time $t_F$ during radiation dominance if 
\begin{align}\label{hrznentryscale}
d_{\rm str}(t_F)=2t_F.
\end{align} 

After horizon re-entry, the strings inter-commute and form loops at any subsequent time $t_i$. These loops of initial length $l_i = \alpha t_i$ decay via emission of gravity waves, and the dominant contribution comes from the largest loops with $\alpha\simeq 0.1$ \cite{blanco}. The gravitational waves can be decomposed into normal modes, and the redshifted frequency of a mode $k$, emitted at time 
$\tilde{t}$, as observed today, is given by \cite{cui}
\begin{align}\label{GWs-freq}
f=\frac{a(\tilde{t})}{a(t_0)}\,\frac{2k}{\alpha t_i - \Gamma G\mu (\tilde{t}-t_i)}, \quad\mbox{with} \quad k = 1,2,3,...
\end{align}
Here $\Gamma\sim 50$ is a numerical factor \cite{vilenkinbook}, $a(t)$ is the scale factor of the universe, and $t_0\simeq 6.62\times 10^{41}~{\rm GeV}^{-1}$ is the present cosmic time. We express the stochastic gravity wave abundance with a present frequency $f$ as 
\begin{align}
\Omega_{\mathrm{GW}}=\frac{1}{\rho_c}\,\frac{d \rho_{\mathrm{GW}}}{d\ln f},
\end{align}
where $\rho_c=3H_0^2m_{\mathrm{Pl}}^2$ is the critical density of the universe, $H_0$ denotes the present value of the Hubble parameter, and $\rho_{\mathrm{GW}}$ is the energy density of the gravity 
waves.
The total gravity wave background coming from all modes is given by \cite{cui}
\begin{align}\label{GWs-Omega}
\Omega_{\mathrm{GW}}(f)=\sum_k \Omega_{\mathrm{GW}}^{(k)}(f),
\end{align}
where 
\begin{align}\label{GWs-Omega-k}
\Omega_{\mathrm{GW}}^{(k)}(f)=&\frac{1}{\rho_c}\,\frac{2k}{f}\,\frac{(0.1)\Gamma k^{-4/3}G\mu^2}{\zeta(4/3)\alpha(\alpha + \Gamma G\mu)}\nonumber \\
& \int_{t_F}^{t_0}d \tilde{t}\,\frac{C_{eff}(t_i)}{{t_i}^4}\left(\frac{a(\tilde{t})}{a(t_0)}\right)^5\left(\frac{a(t_i)}{a(\tilde{t})}\right)^3\theta(t_i-t_F).
\end{align}
Here $\zeta(4/3)=\sum_{m=1}^\infty m^{-4/3}\simeq 3.60$, $C_{eff}(t_i)$ is found to be $0.5$ and $5.7$ for the radiation and matter dominated universe respectively \cite{blanco}, $\theta$ is the Heaviside step function, and the time $t_i$ can be derived from Eq.~(\ref{GWs-freq}).

We calculate the spectrum of gravity waves using Eqs.~(\ref{GWs-Omega}) and (\ref{GWs-Omega-k}). Of course, $\tilde{t}$ should be bigger than $t_i$ since the loops cannot radiate before their formation at $t_i$. The ratios of the type $a(t)/a(t')$ $(t < t')$ are replaced by $(t/t')^{1/2}$ or 
$(t/t')^{2/3}$ if the period between $t$ and $t'$ lies entirely in the radiation or matter dominated period of the universe respectively. If $t < t_{eq} < t'$, where $t_{eq}\simeq 2.25\times 10^{36}~{\rm GeV}^{-1}$ is the equidensity time, i.e. the time at which the radiation and matter energy densities in the universe are equal, we split the time interval between $t$ and $t'$ and express these ratios as 
$(t/t_{eq})^{1/2}(t_{eq}/t')^{2/3}$. Needless to say, the integral in Eq.~(\ref{GWs-Omega-k}) takes into account all the loops that are created at $t_i > t_F$, and radiate a given mode characterized by 
$k$ at $\tilde{t} < t_0$ contributing to the gravity waves with a given frequency $f$. The value of 
$t_i$ is found from Eq.~(\ref{GWs-freq}). Finally, the sum in Eq.~(\ref{GWs-Omega}) is taken over all the modes with $k=1,2,...,10^5$.

\section{Inflation, Gravity Waves and the PPTA Bound}
\label{sec:tension}
\begin{figure}[!htb]
\begin{center}
\includegraphics[width=0.95\linewidth]{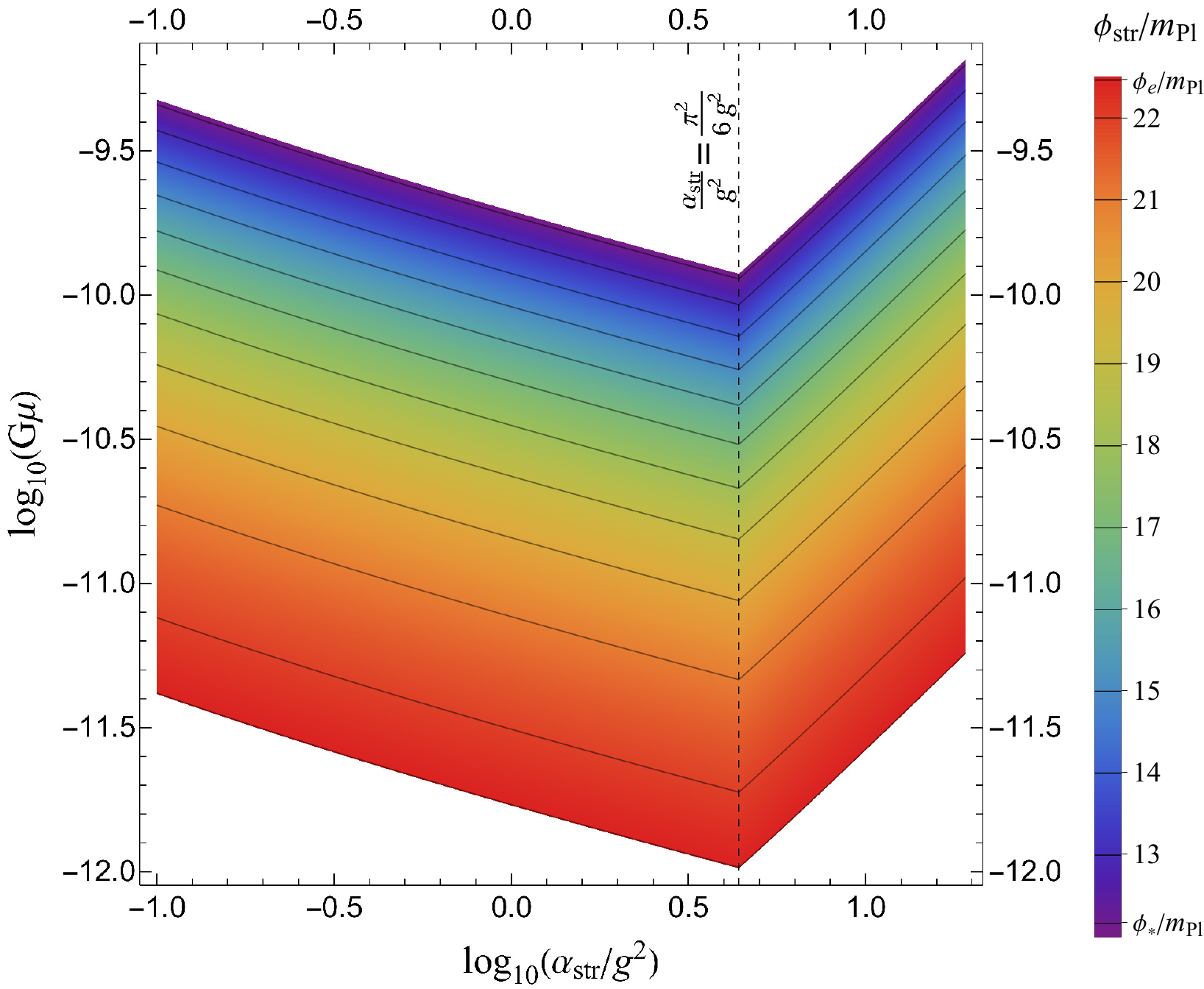}
\end{center}
\caption{The dimensionless string tension $G\mu$ versus $\alpha_{\rm str}/g^2$ for different inflaton values $\phi_{\rm str}$ at the phase transition during which the strings are formed. The rainbow color code shows the variation of $\phi_{\rm str}$. The values of the inflaton field at horizon crossing of the pivot scale ($\phi_*$) and at the end of inflation ($\phi_e$) are also indicated.}\label{plot:gmu_ag_phi}
\end{figure}
\begin{figure}[!htb]
\begin{center}
\includegraphics[width=0.95\linewidth]{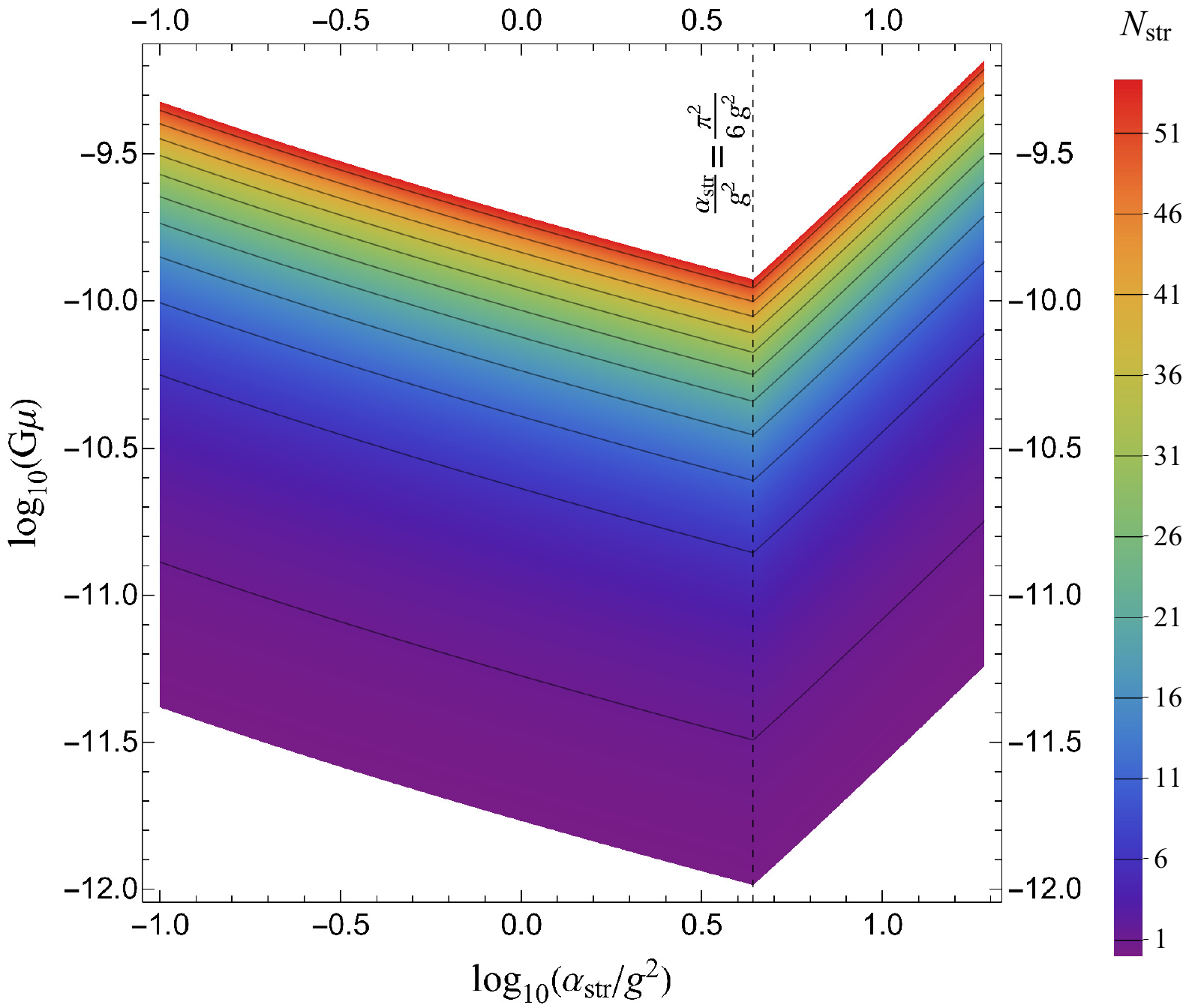}
\end{center}
\caption{The dimensionless string tension $G\mu$ versus $\alpha_{\rm str}/g^2$ for different values of the number of $e$-foldings $N_{\rm str}$ after the phase transition during which the strings are formed. The rainbow color code shows the variation of $N_{\rm str}$.}\label{plot:gmu_ag_N}
\end{figure}
\begin{figure}[!htb]
\begin{center}
\includegraphics[width=0.95\linewidth]{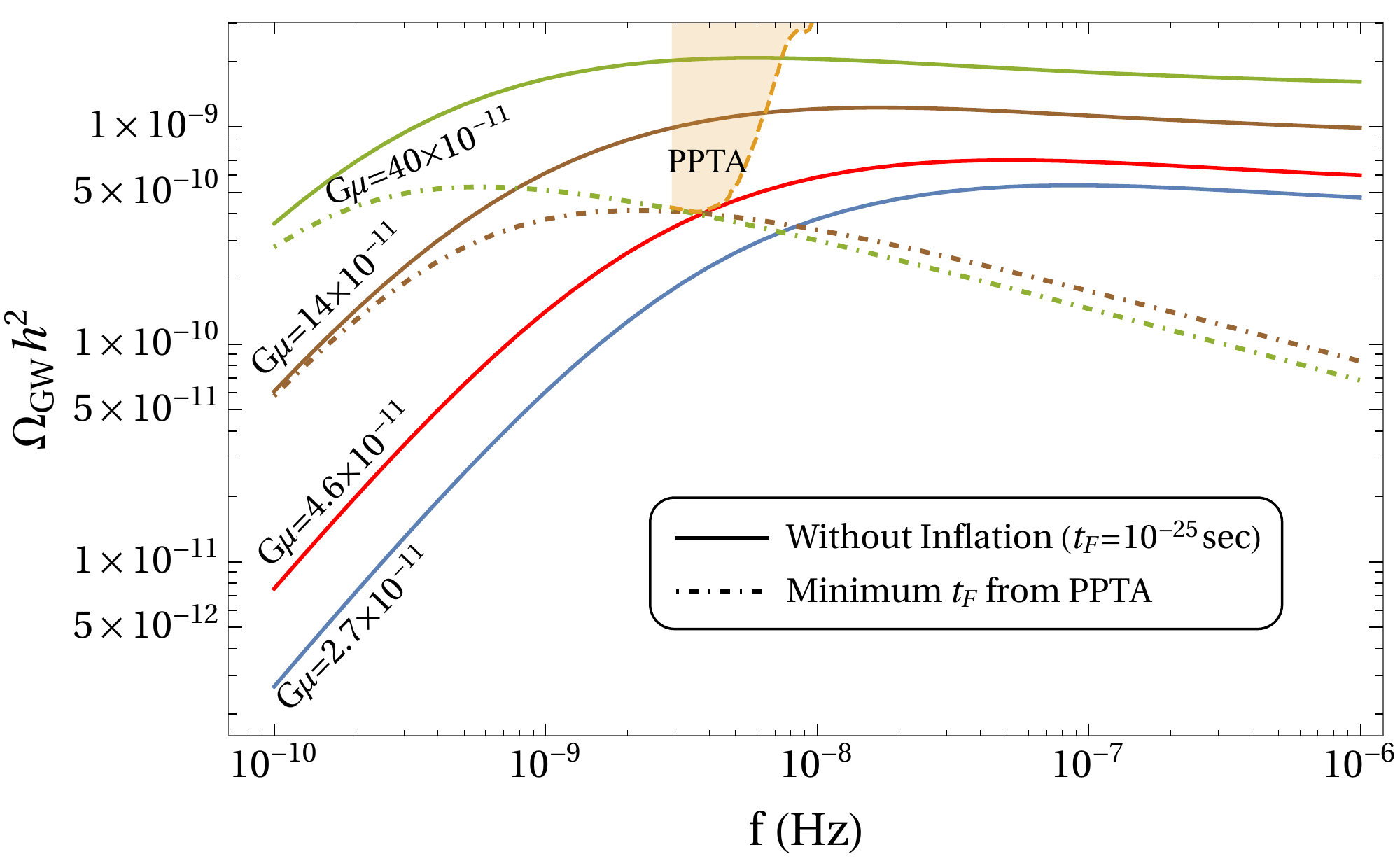}
\end{center}
\caption{Gravity wave spectra with and without inflation for $G\mu=(2.7,4.6,14,40)\times 10^{-11}$ 
(blue, red, brown, and green curves respectively) . Without inflation (solid lines), the PPTA bound 
is violated for $G\mu>4.6\times 10^{-11}$. A minimum horizon re-entry time $t_F$ of the strings is required so that the PPTA bound is satisfied for any given $G\mu$ in this range. The corresponding spectra are denoted by dash-dotted lines.}
\label{plot:GWs-spectra}
\end{figure}
\begin{figure}[!htb]
\begin{center}
\subfloat[$m_{\rm eff}^{-1}\geq H^{-1}$.\label{plot:tF-Gmu-1}]{
\includegraphics[width=0.95\linewidth]{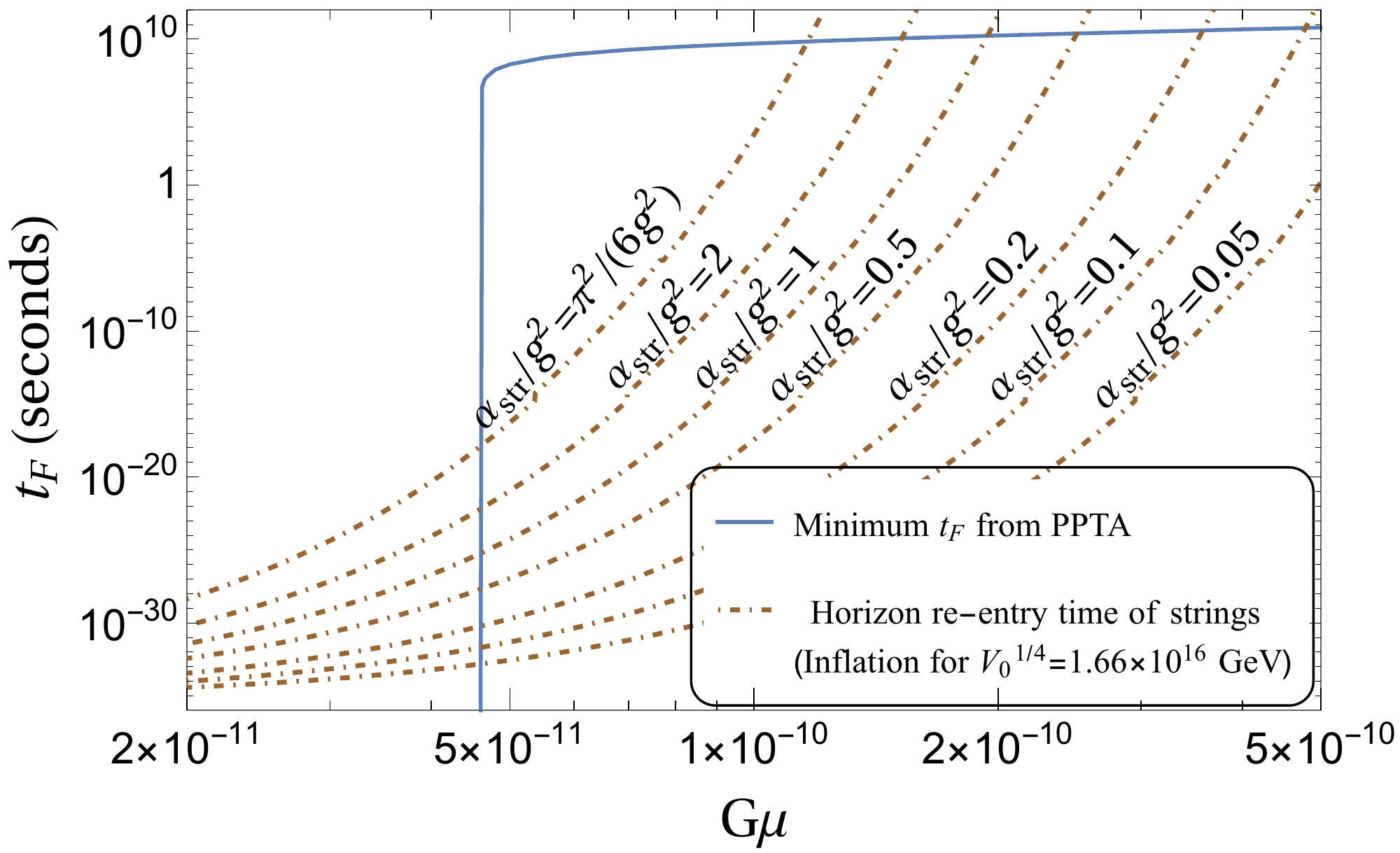}
}\\ 
\vspace*{5mm}
\subfloat[$m_{\rm eff}^{-1}\leq H^{-1}$.\label{plot:tF-Gmu-2}]{
\includegraphics[width=0.95\linewidth]{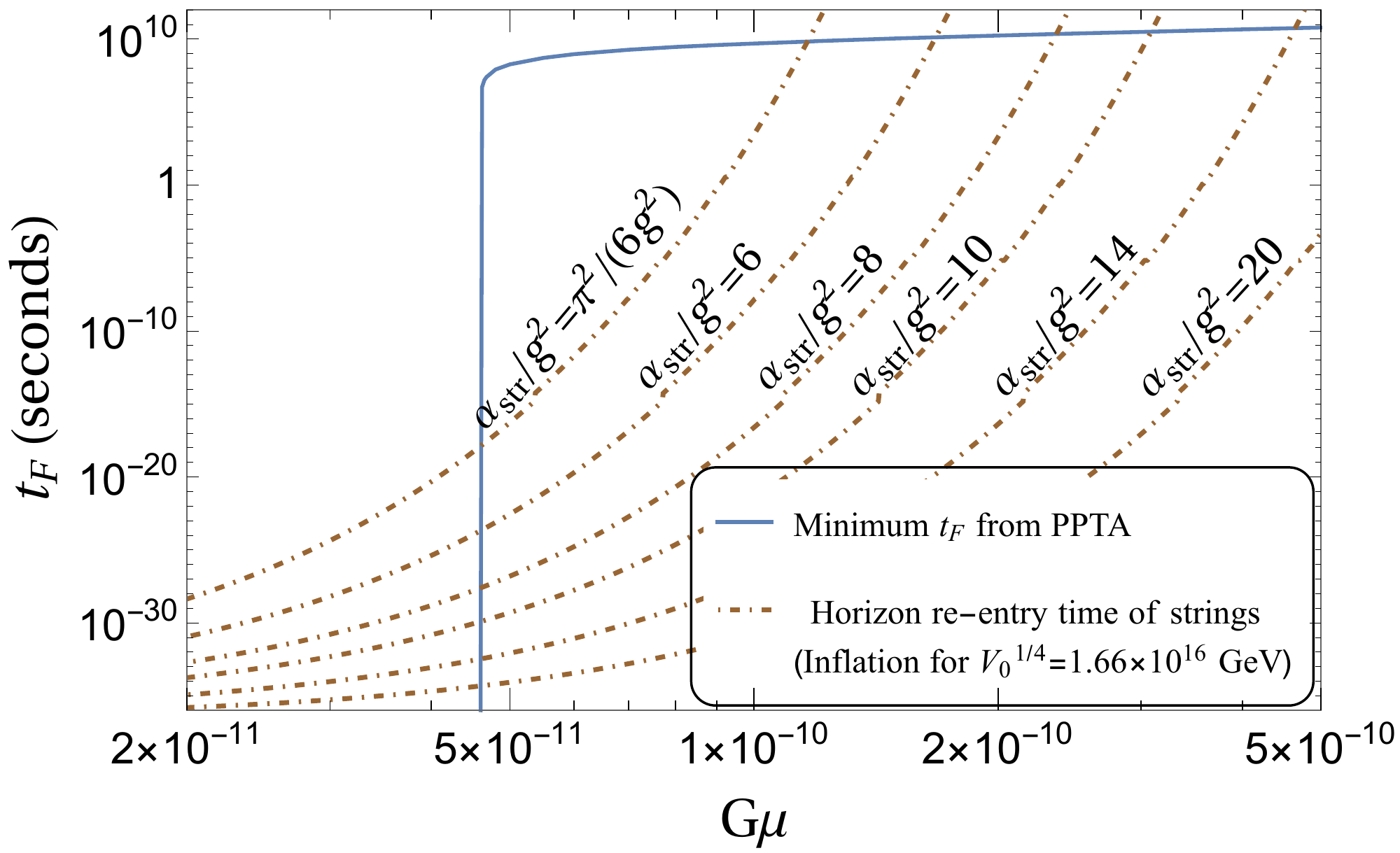}
}
\end{center}
\caption{Horizon re-entry time of the strings as a function of $G\mu$ (brown dash-dotted curves) for different values of the ratio $\alpha_{\mathrm{str}}/g^2$ as indicated. 
 The solid blue lines depict the minimum $t_F$ allowed by the PPTA bound. Notice that for $G\mu\leq 4.6 \times 10^{-11}$, $t_F$ can be very small. Recall that in this paper we assume that the gravity waves originate solely from strings.}
\label{plot:tF-Gmu-zoom}
\end{figure}
\begin{figure}[!htb]
\begin{center}
\includegraphics[width=0.95\linewidth]{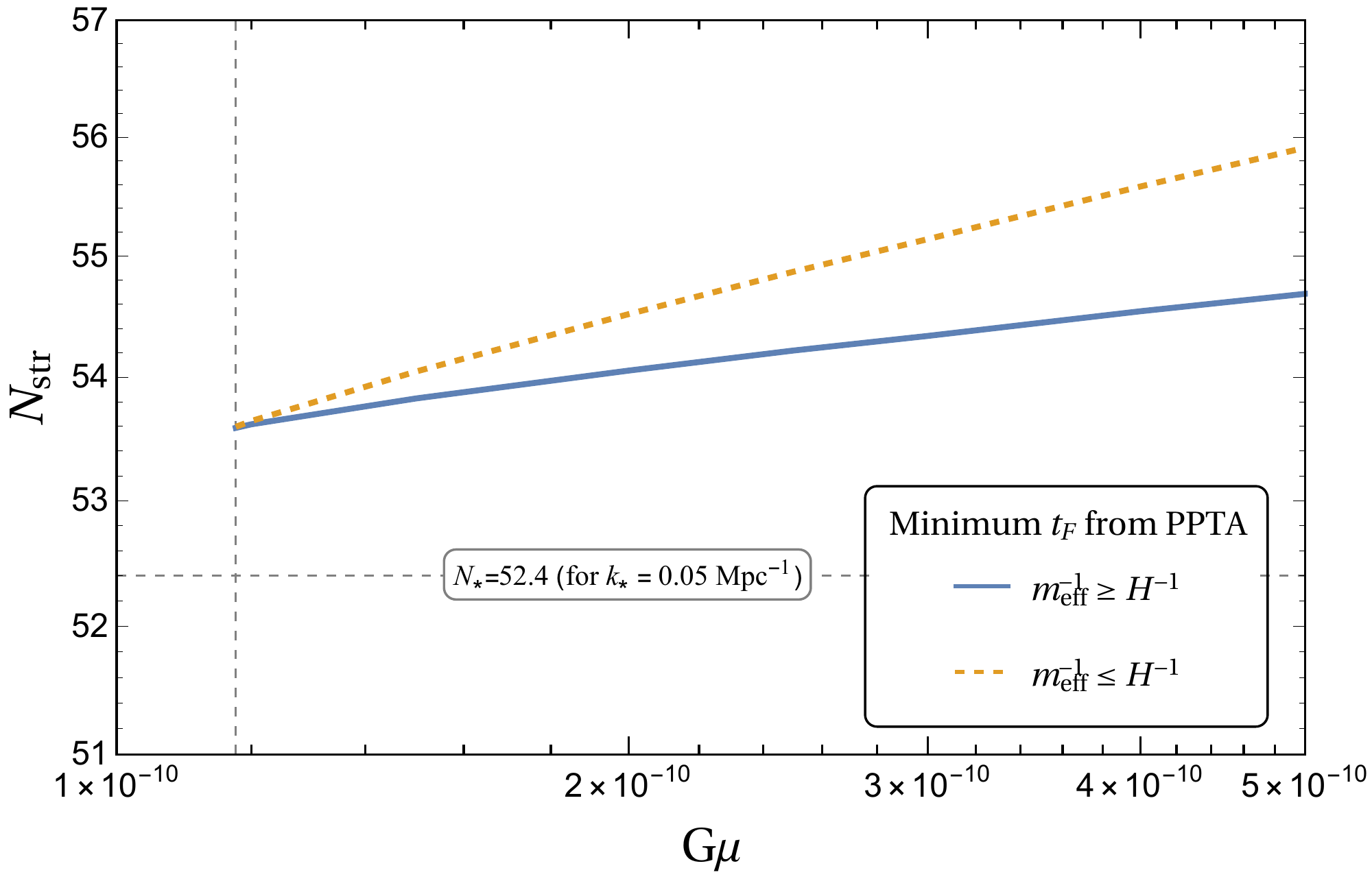}
\end{center}
\caption{The number of $e$-foldings $N_{\rm str}$ experienced by the strings for various $G\mu$ values so that the strings re-enter the horizon at the minimum $t_F$ to satisfy the PPTA bound.}\label{plot:N-Gmu}
\end{figure}
\begin{figure}[!htb]
\begin{center}
\includegraphics[width=0.95\linewidth]{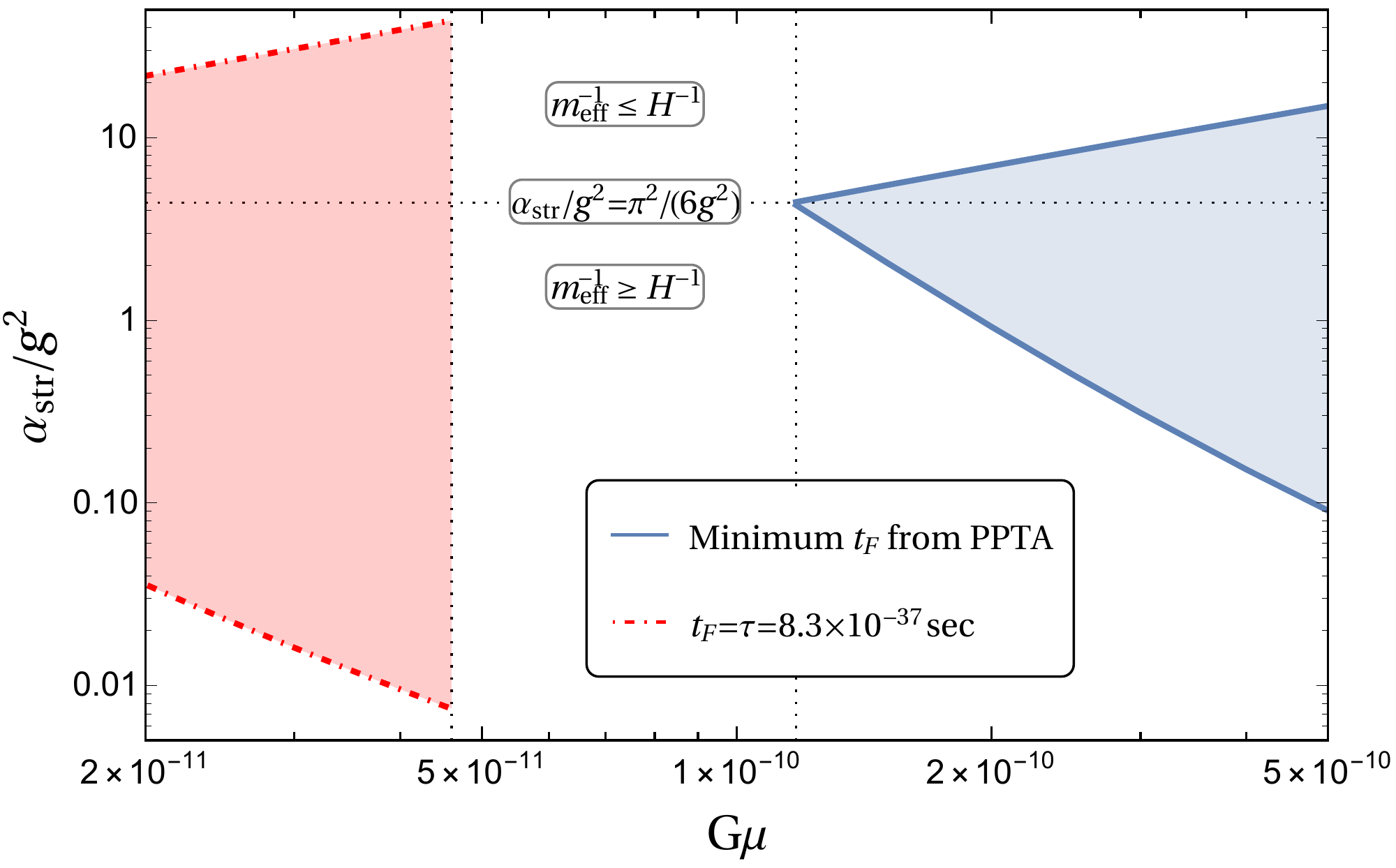}
\end{center}
\caption{Allowed ranges (blue- and red-shaded) of $\alpha_{\mathrm{str}}/g^2$ versus $G\mu$ permitted by PPTA. Blue solid lines are for the minimum $t_F$ permitted by PPTA with $G\mu > 1.1\times 10^{-10}$. For $G\mu \leq 4.6\times 10^{-11}$, the lower and upper bounds on $\alpha_{\mathrm{str}}/g^2$ (red dashed-dotted lines) correspond to $t_F=\tau$. The strings suffer some $e$-foldings in the red-shaded region between them. In the region $4.6\times 10^{-11} < G\mu < 1.1\times 10^{-10}$, there exist no solutions.}\label{plot:ag-Gmu}
\end{figure}
\begin{figure}[!htb]
\begin{center}
\includegraphics[width=0.95\linewidth]{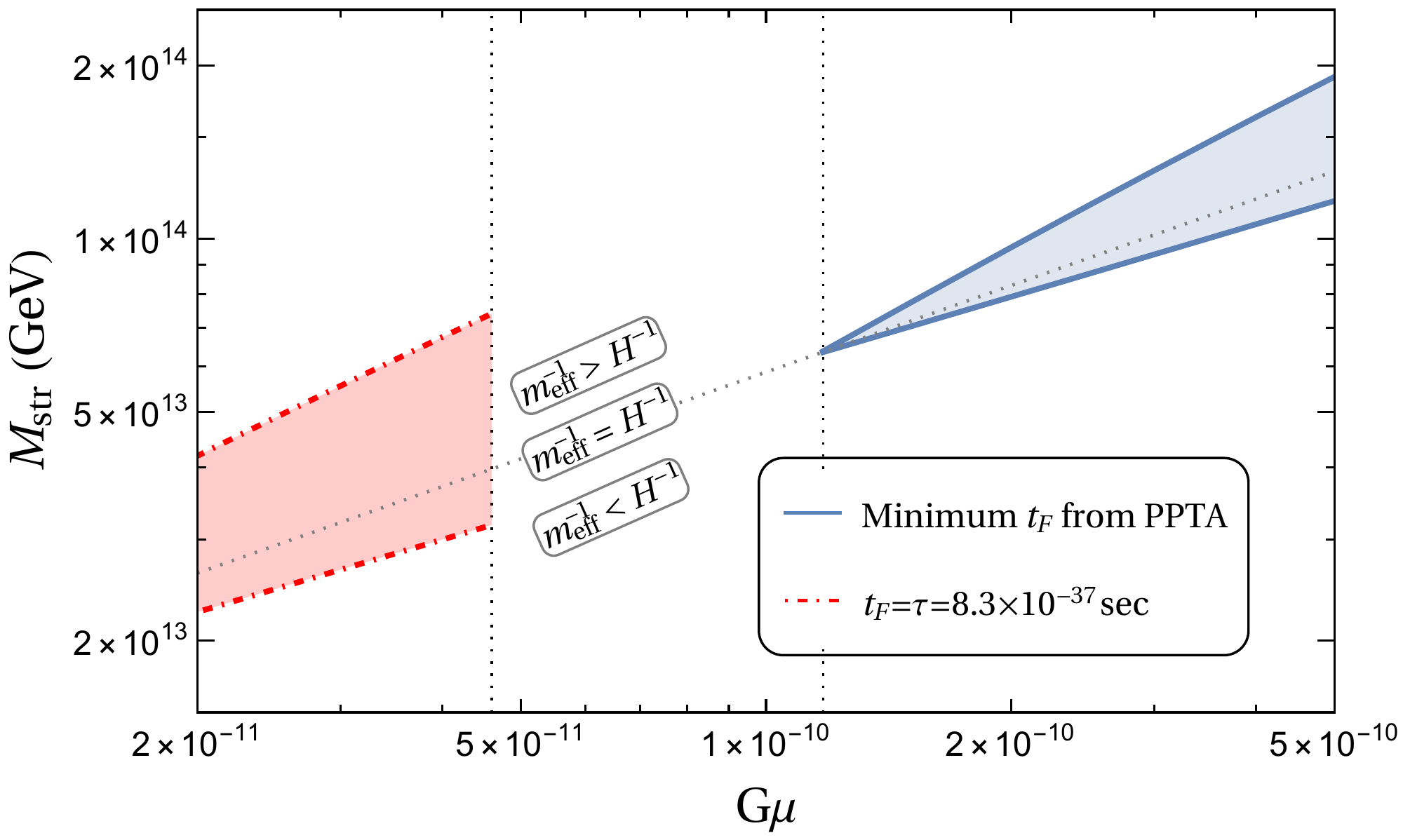}
\end{center}
\caption{ Allowed ranges (blue- and red-shaded) of $M_{\mathrm{str}}$ versus $G\mu$ permitted by PPTA. Blue solid lines are for the minimum $t_F$ permitted by PPTA with $G\mu > 1.1\times 10^{-10}$. For $G\mu \leq 4.6\times 10^{-11}$, the lower and upper bounds on $M_{\mathrm{str}}$ (red dashed-dotted lines) correspond to $t_F=\tau$. The strings suffer some $e$-foldings in the red-shaded region between them. In the region $4.6\times 10^{-11} < G\mu < 1.1\times 10^{-10}$, there exist no solutions for $M_{\mathrm{str}}$.}\label{plot:mII-Gmu}
\end{figure}
We now return to the successful inflationary scenario summarized in Sec.~\ref{sec:inf}. From Eqs.~(\ref{inter-string-dist})-(\ref{hrznentryscale}), we see that the horizon re-entry time $t_F$ of the string network depends on the inflaton field value $\phi_{\rm str}$ at the phase transition during which the strings were generated. On the other hand, $G\mu$ depends on $\phi_{\rm str}$ and $\alpha_{\rm str}/g^2$, as can be deduced from Eqs.~(\ref{breaking_scale_1})-(\ref{Bx}) ($g$ is fixed by the condition of gauge coupling unification). The variation of $G\mu$ with $\alpha_{\rm str}/g^2$ for various values of $\phi_{\rm str}$ and $N_{\rm str}$ is shown in Figs.~\ref{plot:gmu_ag_phi} and \ref{plot:gmu_ag_N}, respectively. Substituting $\phi_{\rm str}$ as a function of $t_F$, we can then find the variation of $t_F$ versus $G\mu$ for different choices of the ratio of the quartic coupling to the gauge coupling squared ($\alpha_{\rm str}/g^2$). We observe from Figs.~\ref{plot:gmu_ag_phi} and \ref{plot:gmu_ag_N} that the contours of $\phi_{\rm str}$ and $N_{\rm str}$ close to $\phi_*$ and $N_*$, respectively, are closely spaced in the region of higher $G\mu$ values ($\gtrsim 10^{-10}$). This implies a slow variation of $G\mu$ with $\phi_{\rm str}$ and $N_{\rm str}$, and hence with $t_F$ in this region (Fig.~\ref{plot:tF-Gmu-zoom}). It is important to emphasize that the new aspect of our model is that the relation between the string formation time and the string tension is different than in the usual scenario.

We compute the stochastic gravity wave spectrum from the cosmic string loops as described in Sec.~\ref{sec:strings}. We first take the time $t_F$ at which the formation of the string loops starts to be very small, which would be the case without inflation. Although the results in this case are insensitive to the precise value of $t_F$, we set $t_F=10^{-25}$ sec for definiteness. The gravity wave spectra for four representative values of $G\mu=(2.7,4.6,14,40)\times 10^{-11}$ are shown in Fig.~\ref{plot:GWs-spectra} by solid lines. We observe that without inflation, in which case $t_F$ is very small, the PPTA bound \cite{ppta} is violated for $G\mu>4.6\times 10^{-11}$. In this case we consider higher values of $t_F$ so that the spectrum satisfies the PPTA bound. If $t_F$ is sufficiently large the smaller string loops will be absent. Consequently, the spectrum will be reduced in the higher frequency regime which can help to satisfy the PPTA bound. In Fig.~\ref{plot:GWs-spectra}, the spectra corresponding to the minimum required value of $t_F$ so that the PPTA bound is satisfied are represented by dashed-dotted lines. 

 With different choices of the ratio $\alpha_{\mathrm{str}}/g^2$ between $0.1$ and $14$, we find the horizon re-entry time $t_F$ of the string network as a function of $G\mu$. The results are displayed in Fig.~\ref{plot:tF-Gmu-zoom} superimposed on top of the minimum allowed values of $t_F$ from PPTA. In Fig.~\ref{plot:tF-Gmu-1}, $\alpha_{\mathrm{str}}\leq \pi^2/6$ and thus the correlation length $\xi=H^{-1}$. In Fig.~\ref{plot:tF-Gmu-2}, on the other hand, $\alpha_{\mathrm{str}}\geq \pi^2/6$ and thus $\xi=m_{\rm eff}^{-1}$. We see that the strings with $G\mu\gtrsim 1.1\times 10^{-10}$ could  suffer sufficient number of $e$-foldings so that the PPTA bound is satisfied within the GUT-inflation
model with the Coleman-Weinberg potential. The number of $e$-foldings $N_{\rm str}$ experienced by the strings are also shown in Fig.~\ref{plot:N-Gmu}. The allowed values of $\alpha_{\mathrm{str}}/g^2$ and $M_{\mathrm{str}}$ for various $G\mu$ values are shown in Figs.~\ref{plot:ag-Gmu} and \ref{plot:mII-Gmu} respectively. These values are compatible with the allowed values of $t_F$ which satisfy the PPTA bound. For $G\mu \leq 4.6\times 10^{-11}$ very small value of $t_F$ are permitted. However, $t_F$ cannot be made smaller than the time $\tau\simeq 8.3\times 10^{-37}~{\rm sec}$ when inflation is terminated. This corresponds to a lower and an upper bound on $\alpha_{\mathrm{str}}/g^2$ in this range depicted in Fig.~\ref{plot:ag-Gmu} by red dashed-dotted lines. The corresponding upper and lower bound on $M_{\mathrm{str}}$ is depicted in Fig.~\ref{plot:mII-Gmu} by red dashed-dotted lines. The strings suffer some $e$-foldings in the red-shaded region. However, the horizon re-entry time remains quite small as can be seen from Fig.~\ref{plot:tF-Gmu-zoom}. The strings above and below this region will be formed after the end of the inflation. 
Note that $N_{\rm str}$ can be arbitrarily small for $G\mu\leq 4.6\times 10^{-11}$, which is consistent with the fact that inflation is not necessary for the PPTA bound to be satisfied in this range. To summarize, we see that Coleman-Weinberg inflation can help to satisfy the PPTA bound for $G\mu$ values in the range $G\mu>1.1\times 10^{-10}$. 

\section{Power-law Approximation for Gravity Waves and NANOGrav}
\label{sec:power}

In the NANOGrav experiment \cite{NANOGrav} the gravity wave spectra are expressed in a power-law form with characteristic strain
\begin{align}\label{GWs-hc-PL}
h_c(f) = A\left( \frac{f}{f_{\mathrm{yr}}}\right)^\alpha,
\end{align}
where $f_{\mathrm{yr}} = 1  \mathrm{yr}^{-1}$ is the reference frequency. This gives
\begin{align}\label{GWs-Omega-PL}
\Omega_{\mathrm{GW}}(f) = \frac{2\pi^2}{3H_0^2}\, f^2h_c(f)^2 = \Omega_0 \left( \frac{f}{f_{\mathrm{yr}}}\right)^{5-\gamma},
\end{align}
where $\Omega_0 = (2\pi^2/3H_0^2)A^2 f_{\mathrm{yr}}^2$ and $\gamma = 3-2\alpha$.
We compute the gravity wave spectrum using Eqs.~(\ref{GWs-Omega}) and (\ref{GWs-Omega-k}) within the frequency range $f\in [2.4,12]\times 10^{-9}$ Hz,
\begin{figure}[htbp]
\begin{center}
\includegraphics[width=0.95\linewidth]{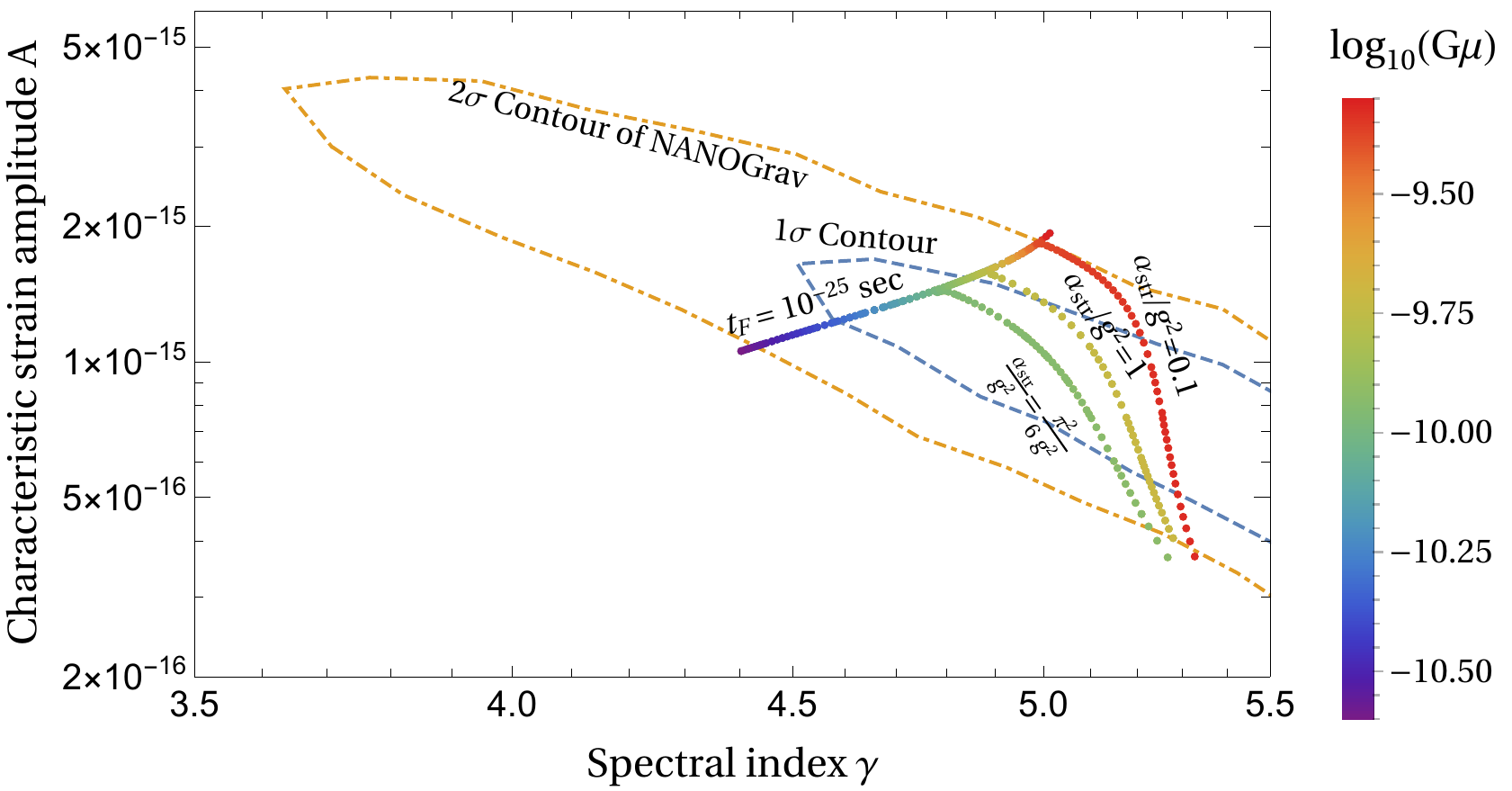}
\par
\vspace{1cm}
\includegraphics[width=0.95\linewidth]{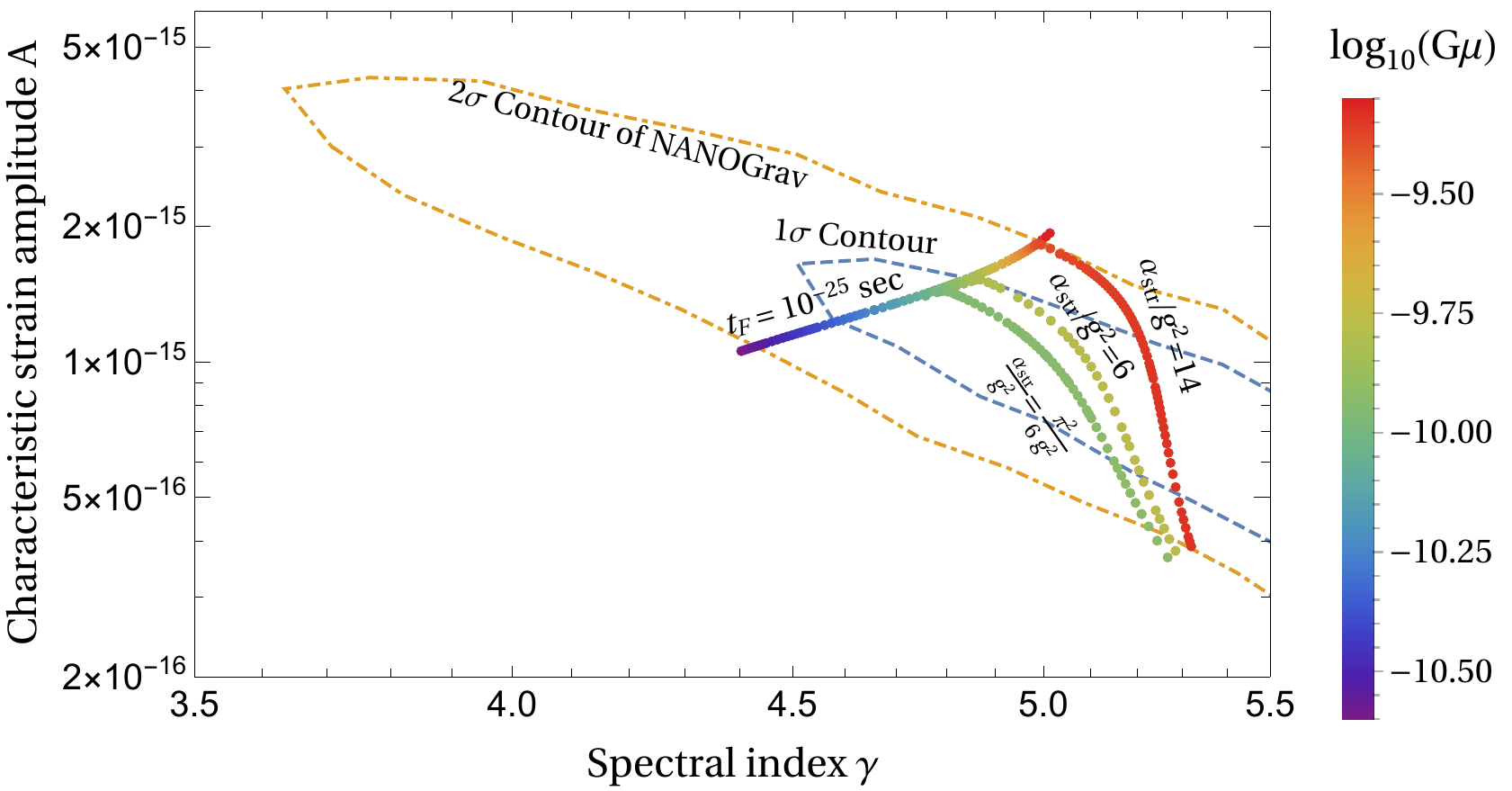}
\end{center}
\caption{The amplitude $A$ of the characteristic strain versus the spectral index $\gamma$ for gravity waves from string loops of different $G\mu$ values is displayed on top of the $1\sigma$ and $2\sigma$ contours of NANOGrav \cite{NANOGrav}. The strings are assumed to: 1. be present in the horizon from a very early time (taken to be $t_F = 10^{-25}$ sec without loss of generality), 2. undergo inflation driven by the Coleman-Weinberg potential of a real GUT singlet with $V_0^{1/4}=1.66\times 10^{16}$ GeV. The plotted values of $A$ and $\gamma$ are for the parameter $\alpha_{\rm str}/g^2$ to be equal to $\pi^2/(6g^2)$, and two other choices $1$ and $0.1$ ($6$ and $14$) for the case $m_{\rm eff}^{-1}> H^{-1}$  ($m_{\rm eff}^{-1}< H^{-1}$) in the upper (lower) panel with a rainbow color code for the variation of $G\mu$. The $1\sigma$ and $2\sigma$ ranges of $G\mu$ satisfying NANOGrav with very low $t_F$ values are $[4.6,14]\times 10^{-11}$ and $[2.7,40]\times 10^{-11}$ respectively.}
\label{plot:A-gamma-1}
\end{figure}
 and fit the results to the power-law expression in Eq.~(\ref{GWs-Omega-PL}) so as to calculate the amplitude $A$ of the characteristic strain and the spectral index $\gamma$. We then compare the calculated $A$ and $\gamma$ values with the NANOGrav $12.5$ yr results \cite{NANOGrav}.

In Fig.~\ref{plot:A-gamma-1}, we show the fitted values of $A$ and $\gamma$ for different values of $G\mu$ with and without inflation. We first take the time $t_F$ at which the formation of the string loops starts to be very small, which would be the case without inflation. Although the results in this case are insensitive to the precise value of $t_F$, we set $t_F=10^{-25}$ sec for definiteness. We find that the NANOGrav 12.5 yr data are well satisfied for $G\mu\in [4.6,14]\times 10^{-11}$ and $G\mu\in [2.7,40]\times 10^{-11}$ within the $1\sigma$ and $2\sigma$ limits respectively. The corresponding gravity wave spectra for $G\mu=(2.7,4.6,14,40)\times 10^{-11}$ are shown in Fig.~\ref{plot:GWs-spectra} by solid lines. 

We then employ the GUT inflation and computed the gravity wave spectra in the relevant frequency range of NANOGrav. The fitted values of $A$ and $\gamma$ are plotted in Fig.~\ref{plot:A-gamma-1} for the parameter $\alpha_{\rm str}/g^2$ to be equal $\pi^2/(6g^2)$, and two other choices $1$ and $0.1$ ($6$ and $14$) for the case $m_{\rm eff}^{-1}> H^{-1}$  ($m_{\rm eff}^{-1}< H^{-1}$) in the upper (lower) panel with a rainbow color code for the variation of $G\mu$. Needless to mention, the gravity wave spectrum remains unaffected in the operating frequency range of NANOGrav unless the horizon re-entry time $t_F$ is sufficiently large. This can be seen from Fig.~\ref{plot:A-gamma-1}, where the fitted $A$ and $\gamma$ values with inflation coincide with those with $t_F = 10^{-25}$ sec for the lower values of $G\mu$ for which $t_F$ is sufficiently small as can be seen from Fig.~\ref{plot:tF-Gmu-zoom}.
\section{Intermediate Mass Monopoles with Strings}
\label{sec:monopole_string}
In this section we will discuss the production of monopoles and strings at the same breaking scale and their subsequent evolution.
\begin{figure}[!htb]
\centering
\includegraphics[width=0.95\linewidth]{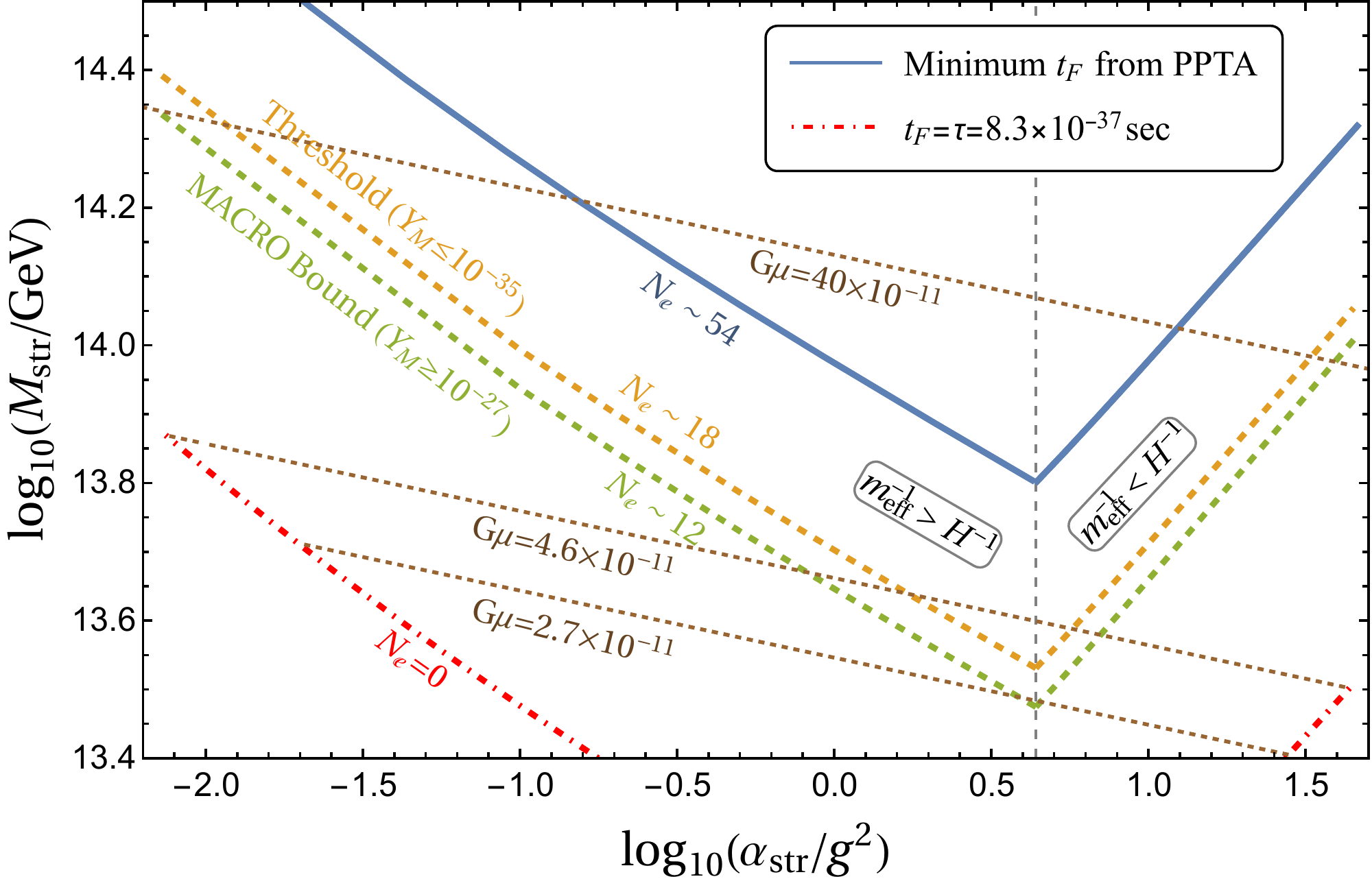}
\caption{$M_{\rm str}$ versus $\alpha_{\rm str}/g^2$ for Coleman-Weinberg inflation with $V_0^{1/4}=1.66\times 10^{16}$ GeV. The solid blue line corresponds to the minimum $t_F$ allowed by the PPTA bound for 
$G\mu > 1.1\times 10^{-10}$. The green dashed line depicts the lower bound on $M_{\rm str}$ corresponding to the upper bound on the monopole flux from the MACRO experiment ($Y_M = 10^{-27}$), and the orange dashed line shows the adopted threshold for observability of the monopole flux ($Y_M = 10^{-35}$). The dashed-dotted red lines correspond to the strings that are created at the end of inflation. The brown dotted lines depict the contours for $G\mu = (2.7,4.6,40)\times 10^{-11}$. $G\mu = (2.7,40)\times 10^{-11}$ correspond to the minimum and maximum string tensions for very small $t_F$ ($\simeq 10^{-25}$ sec) allowed by the NANOGrav $2\sigma$ limit. The minimum allowed $G\mu$ corresponding to the MACRO bound is also about $2.7\times 10^{-11}$. The maximum $G\mu$ value that satisfies the PPTA bound without inflation is $4.6\times 10^{-11}$, which coincidentally lies on the $1\sigma$ contour of NANOGrav.}\label{Fig:string-monopole-region}
\end{figure}
We construct Fig.~\ref{Fig:string-monopole-region} where we plot values of the string scale $M_{\rm str}$ allowed by gravity wave considerations versus $\alpha_{\rm str}/g^2$. Assuming that monopoles are generated at the same scale, we also show the bounds on $M_{\rm str}$ from the monopole flux. The solid blue line corresponds to the minimum $t_F$ allowed by the PPTA bound for $G\mu > 1.1\times 10^{-10}$. The number of $e$-foldings for this line is of the order of $54$. The dashed-dotted red lines at the bottom correspond to the strings appearing at the end of inflation.
The green dashed line represents the upper bound on the magnetic monopole flux from the MACRO experiment \cite{macro} corresponding to a monopole abundance $Y_M =10^{-27}$ \cite{Chakrabortty:2020otp,turnerb}. The orange dashed line, on the other hand, represents the adopted threshold for observability of the monopole flux ($Y_M = 10^{-35}$) \cite{Chakrabortty:2020otp}. The monopole abundance $Y_M$ is calculated using Eq.~(7.1) of Ref.~\cite{Chakrabortty:2020otp} where $m_{\rm eff}^{-1}$ is replaced by the correlation length $\xi$.

If we consider the PPTA bound along with the monopole flux, we see that the strings with $G\mu > 1.1\times 10^{-10}$, which satisfy the PPTA bound, clearly satisfy the MACRO bound too. However, the predicted monopole flux is too low to be observable. The strings with $2.7\times 10^{-11}\lesssim G\mu < 4.6\times 10^{-11}$ that satisfy the PPTA bound can also satisfy the MACRO bound and we can have a measurable monopole flux for some parameter space.

The number of $e$-foldings required to satisfy the MACRO bound is around $12$, and that for the threshold for observability is of the order of $18$. In this range the horizon re-entry time of the strings will be sufficiently small and the NANOGrav data will be satisfied by the strings with $2.7\times 10^{-11}\leq G\mu \leq 40\times 10^{-11}$. Therefore we can have a potentially measurable monopole flux.

\begin{figure}[!htb]
\centering
\includegraphics[width=0.95\linewidth]{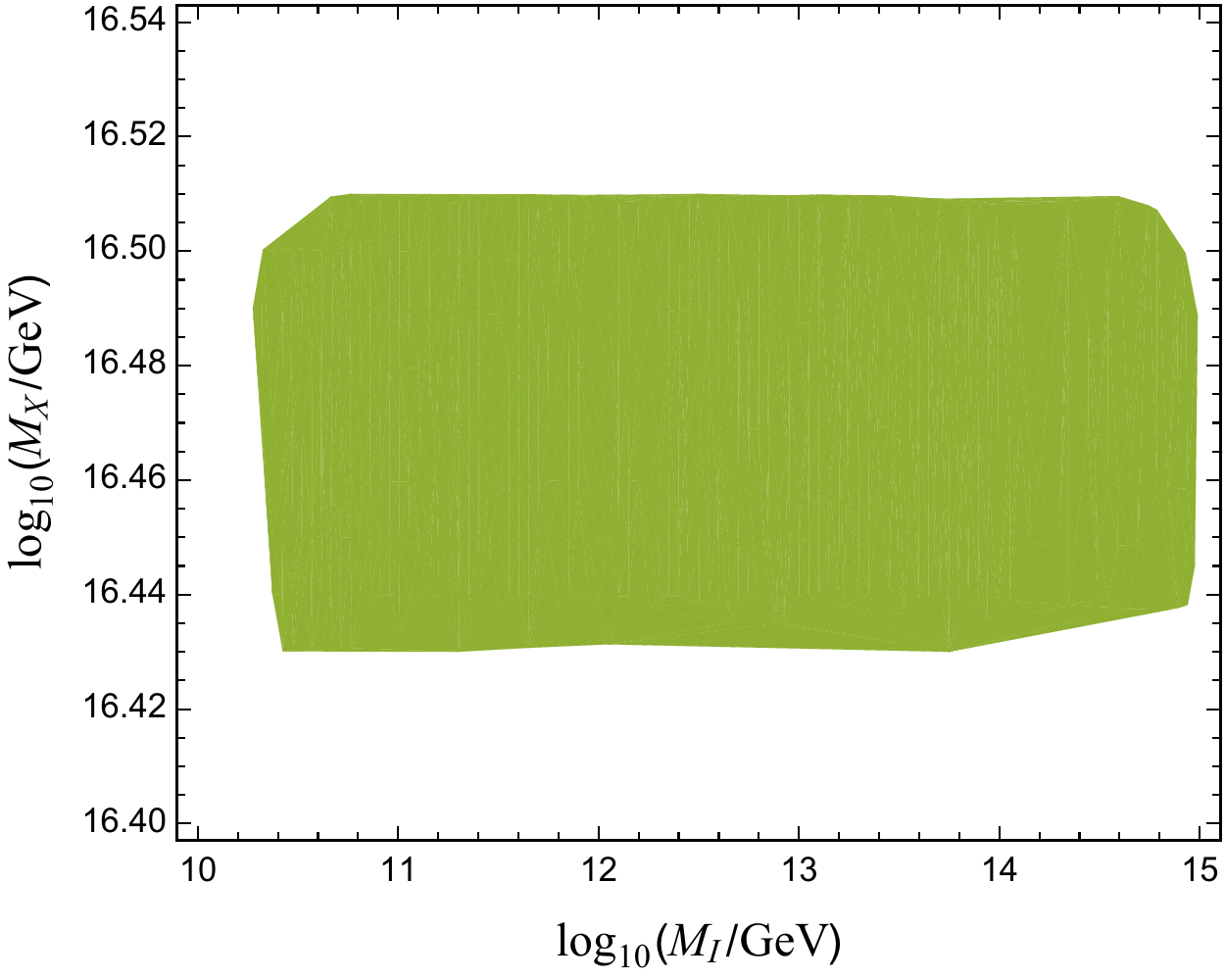}
\caption{The allowed (green) region in the $M_X, M_I$ plane for successful unification and inflation with a Coleman-Weinberg potential for the $E_6$ model with one-intermediate step of symmetry breaking 
(notation as in Ref.~\cite{Chakrabortty:2020otp}). Needless to mention, $M_I\sim 10^{13}-10^{14}$ GeV can incorporate the PPTA bound, MACRO bound, and NANOGrav data.}\label{Fig:E6-G333}
\end{figure}

As an example with one intermediate step of symmetry breaking, we consider $E_6$ broken via the intermediate trinification symmetry $\mathcal{G}_{3_L 3_R 3_C}\equiv SU(3)_L\times 
SU(3)_R \times SU(3)_C$. The VEV of a scalar $650$-plet along the $D$-parity \cite{Dparity} breaking and 
$\mathcal{G}_{3_L 3_R 3_C}$-singlet direction breaks $E_6$ at a scale $M_X$. At this level, a scalar multiplet $(\overline{3},3,1)$ from a $27$-plet and another $(6,\overline{6},1)\subset 351^\prime$ remain massless. The former contains the electroweak Higgs doublet, and the VEV of the latter breaks the trinification symmetry to the standard model. This symmetry breaking produces topologically stable triply charged monopoles and cosmic strings at the same intermediate breaking scale $M_I$ \cite{TopDef,rinku}. The $\beta$-coefficients at the two-loop level that govern the running of the three gauge couplings $g_i$ ($i = 3L, 3R, 3C$) from 
$M_I$ to $M_X$ are given as
\begin{align*} 
b_i = \begin{pmatrix} 
\frac{1}{2} \\ \frac{1}{2} \\ -5
\end{pmatrix} \ ,
\quad b_{ij}= 
\begin{pmatrix}
253 & 220 & 12 \\ 
220 & 253 & 12 \\ 
12 & 12 & 12
\end{pmatrix} \quad \mathrm{with} \quad i = 3L, 3R, 3C.
\end{align*}

We have chosen the heavy gauge boson masses to be equal to the respective breaking scales, and the ratio $R$ of the heavy scalar and fermion masses to the gauge boson masses to vary from $1/4$ to $4$. The solutions for successful unification and inflation with a Coleman-Weinberg potential are shown in Fig.~\ref{Fig:E6-G333} (green area). As we can see, the intermediate scale $M_I$ has solutions in the range $10^{13}-10^{14}$ GeV which can be compatible with the PPTA bound, the MACRO bound, and the NANOGrav data. 

\section{Conclusions} 
\label{sec:concl}

We show how the gravity wave spectrum emitted by intermediate scale 
cosmic strings is affected by primordial inflation. The string network is partially inflated and re-enters the horizon at later times after the termination of inflation. Consequently, the relatively short string loops are not produced which leads to a significant reduction of the gravity wave spectrum at higher frequencies. We consider an inflationary model with a Coleman-Weinberg potential and show how inflation can help to satisfy the PPTA bound for $G\mu$ values in the range $G\mu > 1.1\times 10^{-10}$. We discuss the modification of the gravity wave spectra in the case of Coleman-Weinberg inflation, and identify the spectra which are compatible with the NANOGrav signal. The formation of monopoles and strings at the same intermediate breaking scale is also considered and the compatibility of the MACRO bound on the monopole flux with the PPTA bound or the NANOGrav data is analyzed. Finally, we present a realistic non-supersymmetric $E_6$ GUT model with successful Coleman-Weinberg inflation and one intermediate step of symmetry breaking where both monopoles and cosmic strings are produced. We show that there exists a range of parameters where both monopoles and strings survive inflation and may be present at an observable level.

\section*{Acknowledgments}
This work is supported by the Hellenic Foundation for Research 
and Innovation (H.F.R.I.) under the ``First Call for H.F.R.I. 
Research Projects to support Faculty Members and Researchers and 
the procurement of high-cost research equipment grant'' (Project 
Number: 2251). Q.S. is supported in part by the DOE Grant 
DE-SC-001380. R.M. is supported by the Senior Research Fellowship 
from University Grants Commission, Government of India. 
We thank Joydeep Chakrabortty for his collaboration in the early 
stages of this work.

\def\ijmp#1#2#3{{Int. Jour. Mod. Phys.}
{\bf #1},~#3~(#2)}
\def\plb#1#2#3{{Phys. Lett. B }{\bf #1},~#3~(#2)}
\def\zpc#1#2#3{{Z. Phys. C }{\bf #1},~#3~(#2)}
\def\prl#1#2#3{{Phys. Rev. Lett.}
{\bf #1},~#3~(#2)}
\def\rmp#1#2#3{{Rev. Mod. Phys.}
{\bf #1},~#3~(#2)}
\def\prep#1#2#3{{Phys. Rep. }{\bf #1},~#3~(#2)}
\def\prd#1#2#3{{Phys. Rev. D }{\bf #1},~#3~(#2)}
\def\npb#1#2#3{{Nucl. Phys. }{\bf B#1},~#3~(#2)}
\def\np#1#2#3{{Nucl. Phys. B }{\bf #1},~#3~(#2)}
\def\npps#1#2#3{{Nucl. Phys. B (Proc. Sup.)}
{\bf #1},~#3~(#2)}
\def\mpl#1#2#3{{Mod. Phys. Lett.}
{\bf #1},~#3~(#2)}
\def\arnps#1#2#3{{Annu. Rev. Nucl. Part. Sci.}
{\bf #1},~#3~(#2)}
\def\sjnp#1#2#3{{Sov. J. Nucl. Phys.}
{\bf #1},~#3~(#2)}
\def\jetp#1#2#3{{JETP Lett. }{\bf #1},~#3~(#2)}
\def\app#1#2#3{{Acta Phys. Polon.}
{\bf #1},~#3~(#2)}
\def\rnc#1#2#3{{Riv. Nuovo Cim.}
{\bf #1},~#3~(#2)}
\def\ap#1#2#3{{Ann. Phys. }{\bf #1},~#3~(#2)}
\def\ptp#1#2#3{{Prog. Theor. Phys.}
{\bf #1},~#3~(#2)}
\def\apjl#1#2#3{{Astrophys. J. Lett.}
{\bf #1},~#3~(#2)}
\def\apjs#1#2#3{{Astrophys. J. Suppl.}
{\bf #1},~#3~(#2)}
\def\n#1#2#3{{Nature }{\bf #1},~#3~(#2)}
\def\apj#1#2#3{{Astrophys. J.}
{\bf #1},~#3~(#2)}
\def\anj#1#2#3{{Astron. J. }{\bf #1},~#3~(#2)}
\def\mnras#1#2#3{{MNRAS }{\bf #1},~#3~(#2)}
\def\grg#1#2#3{{Gen. Rel. Grav.}
{\bf #1},~#3~(#2)}
\def\s#1#2#3{{Science }{\bf #1},~#3~(#2)}
\def\baas#1#2#3{{Bull. Am. Astron. Soc.}
{\bf #1},~#3~(#2)}
\def\ibid#1#2#3{{\it ibid. }{\bf #1},~#3~(#2)}
\def\cpc#1#2#3{{Comput. Phys. Commun.}
{\bf #1},~#3~(#2)}
\def\astp#1#2#3{{Astropart. Phys.}
{\bf #1},~#3~(#2)}
\def\epjc#1#2#3{{Eur. Phys. J. C}
{\bf #1},~#3~(#2)}
\def\nima#1#2#3{{Nucl. Instrum. Meth. A}
{\bf #1},~#3~(#2)}
\def\jhep#1#2#3{{J. High Energy Phys.}
{\bf #1},~#3~(#2)}
\def\jcap#1#2#3{{J. Cosmol. Astropart. Phys.}
{\bf #1},~#3~(#2)}
\def\lnp#1#2#3{{Lect. Notes Phys.}
{\bf #1},~#3~(#2)}
\def\jpcs#1#2#3{{J. Phys. Conf. Ser.}
{\bf #1},~#3~(#2)}
\def\aap#1#2#3{{Astron. Astrophys.}
{\bf #1},~#3~(#2)}
\def\mpla#1#2#3{{Mod. Phys. Lett. A}
{\bf #1},~#3~(#2)}

\end{document}